\documentstyle[]{article}
\newcommand{\be}{\begin{equation}}
\newcommand{\ee}{\end{equation}}
\newcommand{\bea}{\begin{eqnarray}}
\newcommand{\eea}{\end{eqnarray}}

\newcommand{\pee}{\mbox{$\mathcal{P}$}}
\newcommand{\gee}{\mbox{$\mathcal{G}$}}

\newcommand{\nn}{\mbox{$\nonumber$}}

\newcommand{\w}{\mbox{$\omega$}}
\newcommand{\de}{\mbox{$\delta$}}
\newcommand{\Ei}{\mbox{$E_{\mbox{\small in}}$}}
\newcommand{\Eo}{\mbox{$E_{\mbox{\small out}}$}}
\begin{document}

\begin{center}
\Large{\bf Dispersion Relations and Relativistic Causality }\\

\vspace{0.25in}

\noindent
{\small  H. Fearn and R. H. Gibb
Department of Physics \\
California State University Fullerton \\
Fullerton CA 92834 USA. \\
phone: (714) 278 2767 \\
fax: (714) 278 5810
email: hfearn@fullerton.edu} \\

\vspace{0.25in}

\noindent
{\bf Abstract}\\

\end{center}

{
In this paper we show that if the refractive index, or rather,
$\left[ n(\w) -1 \right]$ satisfies the
dispersion relations then, it is implied by Titchmarsh's theorem that,
$n(\w) \rightarrow 1$ as $\w \rightarrow \infty$. Any other limiting
value for $n(\w)$ would violate relativistic causality, by which we
mean not only that cause must precede effect but also that signals
cannot travel faster--than--c, the velocity of light in a vacuum. }\\

%\pacs{02.30.Uu, 03.30.+p, 42.25.-p, 42.25.Bs} %RMP does not show pacs!

\noindent
{\small Subj-class: atom-phys; optics}\\


\section{Introduction}

\noindent
There has been much interest in recent theory and experiment on
superluminal
light propagation in media \cite{sci,aep,eric,wang,kuz,mil,segev}.
These articles appear not to violate relativistic
causality, by which we mean the front velocity cannot travel
faster--than--c. The front velocity is the speed at which the very first
extremely small vibrations of the wave will occur \cite{bril}.
The front of the wave is composed of the highest uv frequency components
(ie $\w \rightarrow \infty$). Sommerfeld \cite{somm},
predicted that the front velocity should travel at c. A brief history
of events is given in Brillouin's book \cite{bril}. Later Voigt
\cite{bril}, gave a physical explanation for Sommerfelds result, stating that
the modern theory of dispersion uses the assumption of point like electrons
with a finite mass. Inertia prevents the electrons from oscillating at the very
start of the wave. Electron oscillation needs time to build; only after the wave
has been in motion for some time can the electrons react back on the wave.
The very early oscillations of the wave therefore pass though the dispersive
medium as if through vacuum. Also, electrons having a finite mass, cannot
oscillate at infinite frequencies without amassing infinite energy in the
process. We will restrict our material media to less than
infinite energy.
If $n(\w) \rightarrow \beta$ as $\w \rightarrow \infty$ we take the front
velocity to be $v = c/\beta$, it is
required that $\beta = 1$. In fact we will prove
(via Titchmarsh's theorem to be stated later) that in order that $n(\w)-1$
not violate causality it must obey the dispersion (Kramers Kronig) relations
and as a consequence of this $n(\w) \rightarrow 1$ as
$\w \rightarrow \infty$.

\noindent
This may appear as obvious and well known to some readers. However several well
known text book accounts \cite{jack,sak,panof} on the subject invoke a
physical model for refractive
index to require $n(\w) \rightarrow 1$ at high frequencies as we shall discuss
later. It has not been made clear that the mathematical derivation of the
dispersion relation for $\left[n(\w) -1\right]$ requires
$\left[n(\w) -1\right] \rightarrow 0$ as
$\w \rightarrow \infty$ as part of the proof. The precise physical model is
irrelevant. This is an important point which this article will clarify.

\noindent
It appears not all recent articles agree that
$n(\w) \rightarrow 1$ for high frequency is a requirement for causality
of $n(\w) -1$. It has been claimed that a signal can travel at $v=c/n(\infty)$
where $n(\infty) \leq 1$, \cite{junk}. Furthermore, it was first suggested
that the Scharnhorst effect \cite{scharn,bar,barsh,scharn2},
relating to the normal propagation of light between two parallel mirrors
in a vacuum might allow
a signal velocity (more precisely a front velocity) to exceed c.
In the upcoming section we briefly discuss the Scharnhorst effect
and give arguments which show that the original derivation may be flawed.
Also we show that the effect is extremely small and in practice
unmeasurable. We then concentrate on physical media (non--vacuum) where
experimental verification is possible. Our purpose is to show that
in general signals travel at c for material media despite experimental
claims to the contrary in the literature.
We give physical arguments to show why
$n(\w) \rightarrow 1$ as $ \w \rightarrow \infty$
for these physical media.

\section{Scharnhorst Effect}

\noindent
The Scharnhorst effect relates to light propagating in the vacuum
between two parallel mirrors.
The vacuum modes are changed by the boundaries
(in much the same way as in the Casimir effect) and the light
experiences the vacuum as a dispersive birefringent medium.
The real part of the refractive index in a direction parallel to the
mirror surface is unity. The refractive index in a direction perpendicular
to the mirror surface is found to be less than unity.
These calculations are done using perturbation theory valid only
for small frequencies $\w << m$ where $m$ is the electron mass.\\

Scharnhorst \cite{scharn}, derived a refractive index (perpendicular
to the mirror surface) for the vacuum based on low
frequencies and showed $n(0) <1$. (Note that $n(0)$ implies
the small frequency limit.) Combining this with the Kramers Kronig
relations written for Re $n(\w) -$ Re $n(\infty)$ (as opposed to
Re [$n(\w) -1$]), and setting $\w =0$, Barton and Scharnhorst
\cite{barsh} showed that either (real part) Re $n(\infty) < 1$,
which would imply
signals moving at faster--than--c speed, or (imaginary part)
Im $n(\w) <0$ for some
frequency range, which would imply
that the vacuum could amplify the light signal for some range of frequency.
The paper \cite{barsh}, discusses both options but does not make a definite
choice between the two. \\

The dispersion relations used by Barton and Scharnhorst
\cite{barsh} and later by Scharnhorst alone \cite{scharn2} are not
complete in the sense that they have set a term  Im $n(\infty) = 0$,
which in principle might change their end result.
(See Eq. (\ref{lasteqn}) in this paper). Barton and Scharnhorst say \cite{barsh};

\begin{quote}
``We have no conclusive proof that Re $n$ converges and Im $n$ vanishes as
$\omega \rightarrow \infty$ because [$\cdots$] their asymptotics are quite
likely to be governed by non--perturbative effects that we cannot calculate."
\end{quote}

The reasons for assuming Im $n (\omega \rightarrow \infty )=0$ are given in
their paper \cite{barsh} and we refer the reader there for further discussion.
A further objection to the Scharnhorst effect is that it
violates special relativity (SR). Even if you cannot measure the
effect, to be discussed below,
the fact that it can exist even in principle is objectionable.\\

Consider the case of a light clock, a pair of mirrors with a light
pulse bouncing between them. This type of clock has been used to
derive effects like time dilation and Lorentz contraction in
undergraduate text book accounts of SR. For time dilation the
clock moves in a direction parallel to the surface of the mirrors.
Since the Scharnhorst effect predicts that the vacuum refractive
index is equal to unity in that direction then we would expect the
light clock (in a Scharnhorst type arrangement) to give much the
same prediction as in SR. It should be noted however that the
speed of light in the moving frame of the clock has altered due to
the Scharnhorst effect, so predictions are not exactly the same as
with no Scharnhorst effect. In the Lorentz contraction
arrangement, the light clock is tilted on its side. The direction
of motion of the clock is the same as the light bouncing between
the mirrors, (ie. perpendicular to the mirror surface). It appears
that several light clocks in relative motion would not calculate
the same length scales if the Scharnhorst effect were operating
between the mirrors. Observers in relative motion would not know
what to use for c in the SR velocity addition formula. The change
in the velocity of light, as predicted by the Scharnhorst effect,
is inversely proportional to the fourth power of the distance
between the mirrors, see $\delta c$ below. The distance measured
between the mirrors differs from one reference frame to another in
relative motion due to Lorentz contraction. Hence, the measured
value of the velocity of light, measured from one frame to
another, must also change and this violates SR in a very
fundamental way. The whole of SR is based on c being an invariant.
Which invariant value of c do you use in this case? It appears SR
does indeed break down if the Scharnhorst effect were real, even
if you cannot measure the
Scharnhorst change in c in practice.\\

Two papers have appeared both stating that measurement of
faster--than--c signals between
mirrors was impossible. The first was by Milonni and Svozil \cite{pwm},
which uses an argument based on the uncertainty relation for velocity
and the uncertainty in time due to switching on a signal.
Their argument can be summarized as follows; A measurement of velocity
involves a distance and time measurement $v=L/t$. The time is limited
by a signal turn on time $\delta t \approx 1/\w$ where $\w$ is the
frequency of the signal. We assume the shortest possible delay. Then
$\delta v = L \delta t/ t^2 \geq c^2 \delta t/L$. Using
$c/\w \approx \lambda$ we obtain $\delta v = c \lambda/L$.
The change in the velocity of light predicted by the Scharnhorst effect
is $\delta c = kc \alpha^2 (\lambda_c /L)^4 $ where
$\lambda_c = \hbar/(m c)= 3.9 \times 10^{-11}$cm is the Compton wavelength,
$\alpha = 1/137$ and $k \approx 10^{-2}$.
Hence the ratio of
\be
\frac{\delta v }{\delta c } \geq \frac{1}{k \alpha^2}
\left( \frac{L}{\lambda_c}\right)^3 = 1.5 \times 10^{6}
\left( \frac{L}{\lambda_c}\right)^3
\ee
where we have used $\lambda \approx \lambda_c$. Thus the measured
uncertainty in velocity, $\delta v$, is much greater than the predicted
change in the signal velocity of light, $\delta c$, by many orders of
magnitude. Thus the Scharnhorst effect is not useful as an experimental
verification that faster--than--c speeds for signals are possible.
Milonni and Svozil also conclude \cite{pwm},
\begin{quote}
`` ... it is clear that the uncertainty in the measured propagation
velocity will always be enormously larger than the correction to c associated
with the Scharnhorst effect. We conclude, therefore, that no measurement
of the faster--than--c velocity of light predicted by the Scharnhorst
effect is possible."
\end{quote}
The second paper by Ben--Menahem \cite{ben}, uses an argument based on the
sharpness of the wavefront.
\begin{quote}
``... in order to observe faster--than--c propagation of the wavefront, it
is necessary to sharpen the falloff of the fields at the wavefront to a length
scale less than $1/m$ [where $m$ is the electron mass].
This feat requires the inclusion in the packet of waves
with $\w > m$, for which eq. (3) [the vacuum refractive index derived by
Scharnhorst for $\w \ll m$] is a bad approximation."
\end{quote}

\noindent
In the following sections, we will concentrate on physical media
(non--vacuum media which have only finite energies), in flat space--time,
and we will derive
the dispersion relations for $n(\w) -1$ and in the process prove that
for causal propagation $n(\w) \rightarrow 1$ as $\w \rightarrow \infty$. We
will state Titchmarsh's theorem and use it to prove our result. We
conclude with a discussion that for non--violation of causality it is
a necessary condition that for physical media $n(\w) \rightarrow 1$
for $\w \rightarrow \infty$ and thus signals must travel at c,
the velocity of light in free space.

\section{Pulse propagation through a medium}

\noindent
Consider an electric pulse $E(z,t)$ incident on a thin dielectric medium
of thickness $\de$. The following arguments can be found in the book by
Nussenzveig \cite{nuss}. The pulse is normally incident on the medium and has
a transmission coefficient given by $2/(n(\w) +1)$ which we shall take
to be approximately equal to unity. This turns out to have no influence on
causality considerations. We assume that the medium has
a complex refractive index $n(\w)=n_r (\w) + i n_i (\w) $.
We define the absorption coefficient $\alpha= 2 n_i (\w) \w/c$.\\
\be
E(z,t) = \frac{1}{2 \pi} \int_{-\infty}^{\infty}  \Ei(z,\w) e^{-i\w t} d\w
\ee
The pulse is made up of many Fourier components. The component of
frequency $\w$ can be written as
\be
\Ei (z,\w) = \Ei (0,\w) e^{i\w z/c}
\ee
\noindent
If the pulse travels along the $z$ axis and we place the left face of the
medium at position $z=0$ then
the corresponding electric field component on the exit face of the medium is
\be
\Eo (\de ,\w) = \Ei (0,\w) e^{i\w n(\w)\de/c}
\label{ee}
\ee
\noindent
We will assume that the output is a linear function of the input, so that
the superposition principle applies. This shows that given input $f(t)$
the output $x(t)$ must satisfy
\be
x(t) = \int_{-\infty}^{\infty} g(t,t') f(t') dt'
\ee
\noindent
Furthermore we assume time translation invariance so that if the input is
delayed (or advanced) by a time interval $\tau$ then the output will be
delayed (or advanced) by the same time interval. It follows that
the function $g(t,t')$ must depend only on the time difference between the
input and output pulses, $g(t,t')=g(t-t')$, so that
\be
x(t) = \int_{-\infty}^{\infty} g(t-t') f(t') dt'
\label{conv}
\ee

\noindent
This takes the form of a familiar convolution integral (``Faltung" integral
in German, meaning folding integral). Equation (\ref{conv}) has the
consequence that
\be
X(\w) = G(\w) F(\w)
\label{xgf}
\ee
\noindent
where $F(\w),X(\w)$ and $G(\w)$ are the Fourier transforms of $f(t),x(t)$ and
$g(t)$  respectively. Thus
\be
G(\w) = \int_{-\infty}^{\infty} g(\tau) e^{i\w \tau} d\tau
\ee
If we invoke a primitive causality condition; that the output cannot precede
the input, so that if $f(t)=0$ for $t<\tau$ then the same is true for $x(t)$.
Hence $g(\tau)=0$ for $\tau < 0$ and we can replace the lower bound in the
above integral by zero.
\be
G(\w) = \int_0^{\infty} g(\tau) e^{i\w \tau} d\tau
\label{gofw}
\ee
This implies that $G(\w)$ has a regular analytic continuation over the upper
half of the complex plane, $I_+$. If we write $\w = x+iy$ for $y\geq0$ then
\be
G(\w) = \int_0^{\infty} g(\tau) e^{i x\tau} e^{-y \tau} d\tau
\label{par}
\ee
the decay factor $\exp(-y\tau)$ can only help convergence when $\tau \geq0$.
We still do not have sufficient conditions to derive the dispersion relations
for $G(\w)$. We need information regarding the limiting form of $G(\w)$ as
$\w \rightarrow \infty$. In particular, we need to prove that it decreases
rapidly at infinity, otherwise the Re $G(\w)$ and Im $G(\w)$ may be
totally unrelated as in the case of a complex constant a+ib, \cite{nuss}.
We take $G(\w)$ to be square integrable along the real axis,
(in mathematical notation, $G(\w) \in L^2(-\infty,\infty)$), so that
\be
\int_{-\infty}^{\infty} | G(\w)|^2 d\w < K
\ee
\noindent
where $K$ is a finite constant. If $F(\w)$ and $G(\w)$ are Fourier Transforms
of $f(t)$ and $g(t)$ defined in a similar fashion to
$\Ei(t)$ then according to Parseval's theorem
\be
\int_{-\infty}^{\infty} f(t) g^{\star}(t) dt = \frac{1}{2\pi}
\int_{-\infty}^{\infty} F(\w) G^{\star}(\w) d\w \;\;.
\ee
\noindent
Since both $f(t)$ and $g(t)=0$ for $t<0$, when we set $f(t)=g(t)$ in
the above integral equation we get,
\bea
\int_0^{\infty} |g(t)|^2 dt &=& \frac{1}{2\pi} \int_{-\infty}^{\infty}
|G(\w)|^2 d\w \nn \\
&\leq& \frac{K}{2 \pi}
\label{happy}
\eea
\noindent
Also, since $G(x+iy)$ is the Fourier Transform of $\exp(-yt)g(t)$ from
Eq.(\ref{par}), then by Parseval's theorem
\bea
\int_{-\infty}^{\infty} |G(x+iy)|^2 dx &=& 2\pi \int_0^{\infty}
e^{-2yt} |g(t)|^2 dt \nn \\
&\leq& 2 \pi \int_0^{\infty} |g(t)|^2 dt
\mbox{\hspace{0.5in} ($y\geq0$)} \nn \\
&\leq& K \mbox{\hspace{1.3in} ($y\geq0$)}
\label{axis}
\eea
\noindent
where Eq.(\ref{happy}) has been used in the last step. Equation (\ref{axis}),
expresses the property that if a causal transform is square integrable along
the real axis, then once it is analytically continued into $I_+$, it is
also square integrable along any line parallel to the real axis for
$0 \leq y \leq \infty$. This is a very important result and leads to
another condition that $|G(x+iy)| \rightarrow 0$ for $y\geq0$ when
$x\rightarrow \pm \infty$. This also forms part of Titchmarsh's Theorem 93,
Lemma \cite{titch}, which we shall define in the upcoming section.\\

\noindent
Now to apply these results to our situation, first we note that we are dealing
with electric fields whose intensity must be finite, hence both the input
and output can be taken to be square integrable. Mathematically,
\be
\int_{-\infty}^{\infty} | E(\w) |^2 d\w < \infty
\ee
\noindent
which holds for both the input $\Ei$ and output $\Eo$.
Considering the frequency dependent terms, we note that, Eq.(\ref{ee})
is of the same form as Eq.(\ref{xgf}). Here $X(\w) \equiv \Eo (\de,\w)$,
$F(\w) \equiv \Ei (0,\w)$ and
\be
G(\w) \equiv  e^{i\w n(\w)\de/c}
\label{gt}
\ee
\noindent
Hence
\be
g(\de,\tau) = \frac{1}{2 \pi} \int_{-\infty}^{\infty}
e^{i\w n(\w)\de/c} e^{-i \w \tau} d\w
\ee
where $\tau = t-t'$ and
\be
\Eo (\de ,t) =  \int_{-\infty}^{\infty} g(\de,t-t') \Ei (0,t') dt'
\ee
The integral for $G(\w)$ is defined for negative and well as positive
frequency. The refractive index is usually defined for positive frequency
only. Since $\Ei$ and $\Eo$ are both real, $g$ must be real hence
\be
n(-\w) = n^{\star}(\w)
\label{sym}
\ee
\noindent
for real $\w$ which extends the definition of
$n(\w)$ for the negative frequencies.
Due to the spatial separation $\de$ between input and output
we need to invoke the relativistic causality condition rather than the
``primitive" causality condition quoted earlier. We need to restrict out output
signal so that no signal can propagate with velocity greater than the velocity
of light, $c$. This implies that $\Eo (\de,t)$ can only depend on $\Ei (0,t')$
when $t' < t - \de/c$, since the output time $t = t' + \de/c$. The signal
takes a time $\de/c$ to traverse the medium. Hence $g(\de,\tau) =0$ when
$\tau < \de/c$. Furthermore, from Eq.(\ref{gt}),
\be
G(\w) = e^{i\w n(\w)\de/c} = \int_{\de/c}^{\infty} g(\de,\tau)
e^{i\w \tau} d\tau
\ee
Note that usually $|G(\w)|\leq 1$ even without the transmission coefficient
since the complex refractive index results in exponential decay.
It is usually the case that the output signal intensity is less than or
equal to the input signal intensity. The above equation can be written as,
\be
G(\w) = e^{i\w \left[n(\w)-1 \right] \de/c} = \int_0^{\infty} g(\de,t' + \de/c)
e^{i\w t'} dt'
\label{anal}
\ee
\noindent
The right hand side of this equation has a regular analytic continuation
in $I_+$ (the upper half of the complex plane),
this implies that the left hand side
of the equation does also \cite{nuss,titch}.
Titchmarsh \cite{titch}, gives a very clear criterion for analytic continuation
of a function which is zero for all negative values of its argument. His
theorem 93 has been restated by Toll \cite{toll} as;

\begin{quote}
``A function of integrable square is zero for all negative values of its
argument if and only if its Fourier transform is a causal transform".\\
\end{quote}

\noindent
Thus the causality condition that $g(\de,t' + \de/c)=0$ for negative $t'$
is equivalent to the requirement that $G(\w)$ be a causal transform.

%----------------------------------------------------------
\section{Titchmarsh's Theorem}

\noindent
Theorems 93--95 of Titchmarsh can be restated as follows:

\noindent
Let $\gee (\w)$ be an analytic and square integrable function along the
real axis. (ie. $\gee (\w) \in  L^2(-\infty,\infty)$).  Then if $\gee (\w)$
obeys one of the three conditions below, it obeys all three of them.

\noindent
(i) No output before the input, causality condition:
If $g(\tau)$ is the inverse Fourier transform of $\gee (\w)$, then
\be
g(\tau) =0 \mbox{ \hspace{0.5in} ($\tau<0$) }
\ee
\noindent
this means $\gee (\w)$ is a causal transform.

\noindent
(ii) $\gee (x)$ is, for ``almost all" $x$, the limit as $y \rightarrow 0$ of an
analytic function $\gee (x+iy)$ that is regular (holomorphic) in $I_+$
(the upper half of the complex plane) and square integrable over any line
parallel to the real axis. Mathematically this means,
\be
\int_{-\infty}^{\infty} |\gee (x+iy)|^2 dy < K \mbox{ \hspace{0.5in} $(y\geq0)$}
\ee
where $K$ is a finite real constant. See our derivation of Eq.(\ref{axis})
above. It follows that
\be
\lim_{\mbox{\small $x \rightarrow \pm \infty$}} \gee (x+iy) = 0
\mbox{\hspace{0.5in} $(y \geq 0)$}
\ee
\noindent
The above limiting result comes from the following consideration;
Cauchy's theorem for $y \geq 0$ gives,
\bea
\gee (\w) &=& \frac{1}{2 \pi i} \int_{\Gamma}
\frac{\gee (\nu) d\nu}{\nu -\w} \;\;
\mbox{\hspace{0.5in} ($\w, \;\nu$ complex)} \nn \\
&\equiv& \frac{1}{2 \pi i} \int_{-\infty}^{\infty}
\frac{ \gee (\nu) d\nu}{\nu - \w}
\eea
\noindent
Knowing only that $\gee (\w)$ is
square integrable is {\em not} sufficient to imply causality.
(Nussenzveig \cite{nuss}, gives a counter example in his book page 28.)
A causal transform which is square integrable along the real axis is
also square integrable about any line parallel to the real axis
in the upper half of the complex plane.
Consider a square contour $\Gamma$ in the use of Cauchy's theorem above,
see Titchmarsh pp 125--128 \cite{titch} Theorem 93, Lemma. (See also
the treatment by Nussenzveig pages 22--24 \cite{nuss}).
The integrals along the vertical
sides must tend to zero as $x \rightarrow \pm \infty$. This calculation is
repeated in the Appendix for completeness.

\noindent
(iii) Re \gee (\w) and Im \gee (\w) are conjugate functions
and as such are Hilbert
transforms of each other. ie. $\gee (\w_r) = \gee_r(\w_r) + i \gee_i(\w_r)$ is
a causal transform if and only if its real and imaginary parts satisfy;
\bea
\gee_r(\w_r) &=& \frac{P}{\pi} \int_{-\infty}^{\infty}
\frac{ \gee_i(\nu) d\nu}{\nu - \w_r} \;\;, \nn \\
\gee_i(\w_r) &=& -\frac{P}{\pi} \int_{-\infty}^{\infty}
\frac{ \gee_r(\nu) d\nu}{\nu - \w_r} \;\; .\nn \\
\eea
\noindent
where the frequency is treated as real $\w = \w_r$.
Each of the above expressions implies the other. Here $P$ indicates the
principal part of the integral, (sometimes denoted by a bar through the
integral sign) to be evaluated at $\nu = \w_r$.
These relations between the real and imaginary parts of $\gee (\w)$ are the
dispersion relations also known as Plemelj formulas.
%----------------------------------------------------------

\section{Dispersion Relations}

\noindent
Returning to our specific example, and Eq.(\ref{anal}),
since $\de$ is very small, but otherwise arbitrary,
$\w \left[ n(\w) -1 \right] $ must have the same
property (as $G(\w)$) of analytic continuation in $I_+$ \cite{nuss}.
Nussenzveig also shows that the refractive index is regular in the
upper half of the complex plane. However, this does not exclude the possibility
of singularities on the real axis. This condition is also not sufficient to
show that the real and imaginary parts of $\left[ n(\w) -1 \right] $
must be conjugate functions, or equivalently, that they obey dispersion
relations and hence imply causality.  We must also show that
$ \left[ n(\w) -1 \right] \rightarrow 0 $ as $\w \rightarrow \infty$, which
is part of condition (ii) in Titchmarsh's theorem above. This is in general not
easy to do without resorting to a physical model for the refractive index.

\subsection{Using a Model for refractive index}

\noindent
The usual approach in the literature, taken by both Nussenzveig \cite{nuss} and
several text books \cite{jack,sak,panof}, is to assume that for
very high driving frequency, the frequency
will be much higher than any binding frequency of the electrons in the
medium, and so the electrons will behave as though they are free. The
Lorentz equation of motion becomes
\be
m \ddot{x} = e E(\w) e^{-i\w t}
\ee
\noindent
which would involve a trial solution of the form $x(t) = x(\w) \exp(-i\w t)$
and hence a result $x(\w ) = -e E(\w )/( m \w^2)$. Now since the polarization
of the medium is given by
\be
\pee(\w) = N e x(\w) = \chi(\w) E(\w)
\label{polariz}
\ee
\noindent
where $N$ is the number density of the electrons, $m$ is the electron mass,
and in Gaussian units, the electric susceptibility $\chi(\w)$ is,
\be
\chi(\w) = \frac{1}{4 \pi} \left[ n^2(\w) -1 \right] = -\frac{ N e^2}{m \w^2}
\mbox{ \hspace{0.5in} $(\w \rightarrow \infty)$}
\label{xofw}
\ee
\noindent
Hence, $n(\w) \rightarrow 1$ as $\w \rightarrow \infty$ from physical
considerations. This shows that $\left[ n^2(\w) -1 \right] \in
L^2(-\infty,\infty)$, is
square integrable along the real axis. Here, $\chi(\w)$ plays the role of
$G(\w)$. The causality condition, that you get no polarization before the
electric field is applied, implies that $\chi(\w) =0$ for $t<0$, so that
condition (i) of Titchmarsh's theorem is satisfied. Hence,
\be
\int_{-\infty}^{\infty} |n^2(x+iy)-1|^2 dy < K \mbox{\hspace{0.5in}
$(y\geq 0)$}
\label{one}
\ee
\noindent
where $K$ is a finite real constant.
Thus, from physical considerations,
$\left[ n(x+iy)-1 \right]$, which
is analytic in $I_+$, from the above discussion, must approach zero
as $x\rightarrow \pm \infty$ when $y\geq 0$. This implies,
\be
\int_{-\infty}^{\infty} | n(x+iy) -1|^2 dy < K' \mbox{\hspace{0.5in}
$(y\geq 0)$}
\label{two}
\ee
\noindent
where $K'$ is a real, finite constant.
Since this inequality satisfies condition (ii), then $\left[ n(\w) -1\right]$
must also satisfy the dispersion relations and be a causal transform.

\subsection{No physical model for refractive index}

\noindent
We have yet to make clear that the physical model is not
essential to the derivation of the dispersion relations. We shall now
obtain the same results as in section 5.1 without invoking the physical model
for refractive index. We begin by considering the polarization Eq.
(\ref{polariz}). This is clearly of the form Eq. (\ref{xgf}), where the
polarization $\pee (\w) \equiv X(\w)$, $E(\w) \equiv F(\w)$ and
$\chi(\w) \equiv G(\w)$. The intensities of the electric fields must remain
finite so it is reasonable to assume these quantities are square integrable
along the real axis.
$\chi(\w)$ may also by square integrable or it may satisfy the weaker
restriction of being bounded $|\chi(\w)|^2 \leq K_0$, where $K_0$ is a constant.
In this case it is possible to form a new function by the method of
 subtraction (see discussions section 6.) which is square integrable.
For physical media (of finite energy)
it is reasonable to assume that the function $\chi(\w)$
is decreasing and that it tends to zero for high frequency in all
physically realizable situations.  Electrons in a real media cannot
oscillate at infinite frequency which requires infinite energy and thus
the medium cannot become polarized at infinite frequency.
Furthermore, the causality condition, that you get no
polarization before the electric field impinges on the medium,
requires that $\chi(\w)$ is also causal.
Since $\chi(\w)$ is proportional to $n^2(\w) -1$
(see Eq. (\ref{xofw}) ) then $n^2(\w) -1$ is causal, and also
square integrable. Now it appears that $n^2(\w) -1$ satisfies Titchmarsh's
condition (i), so that implies it satisfies all conditions. We can therefore
write down a dispersion relation for $n^2(\w) -1$ and in doing so we
assume that (by condition (ii) of Tichmarsh's theorem)
$n^2(\w) -1 \rightarrow 0$ as $\w \rightarrow \infty$. This in turn implies
that $n(\w) \rightarrow 1$ as $ \w \rightarrow \infty$ note without
having invoked any physical model for refractive index.

\noindent
Returning to our earlier discussion, in section 3. and at the beginning
of section 5., we had found that $n(\w) -1$ could be extended into $I_+$.
We may assume that $n(\w) -1$ would be
at least bounded. The remaining requirement was
that $n(\w) -1 \rightarrow 0$ as $\w \rightarrow \infty$ in
order to satisfy Titchmarsh's condition (ii).
In considering $n^2(\w) -1$ we have found that
indeed $n(\w) \rightarrow 1$ as $\w \rightarrow \infty$. Now we have
satisfied condition (ii) of Titchmarsh's theorem for $n(\w) -1$ and thus
we may write down the dispersion relations for $n(\w) -1$.

\subsection{Calculation of the Dispersion Relations}

\noindent
Now that we have established that $n(\w) -1$ satisfies Titchmarsh's theorem,
whether we invoke a physical model for refractive index or not,
we may calculate the dispersion relation.
Cauchy's theorem expresses $ \left[ n(\w) -1 \right]$ at any point
in $I_+$, in terms of a contour integral, for example
a large semi--circle in the upper half plane.
Here we define frequency as a complex quantity $\w = \w_r + i \w_i$.
Since we have shown that the refractive index (for real frequency)
$n(\w_r) \rightarrow 1$ as $\w_r \rightarrow \infty$ the contour integral
reduces to an integral over the real axis, as indicated by condition (ii) in
Titchmarsh's theorem above, and derived for a rectangular integral in the
Appendix.  Hence we have,
\be
\left[ n(\w) -1 \right] = \frac{1}{2 \pi i} \int_{-\infty}^{\infty}
\frac{ \left[ n(\nu)-1\right] d\nu}{\nu - \w}
\mbox{\hspace{0.5in} (Im $\w \geq 0$ )}\;\; .
\ee
Taking the limit as $\w$ approaches the real axis, we set complex $\w
\rightarrow \w_r + i\epsilon$
where $\epsilon$ is a small term, the radius of a small semi--circle in the
upper half plane about the real frequency $\w_r$. We shall assume that
$\w_r$ represents a pole on the real axis.
We obtain,
\be
\left[ n(\w) -1 \right] = \frac{1}{2 \pi i} \int_{-\infty}^{\infty}
\frac{ \left[ n(\nu)-1\right] d\nu}{\nu - \w_r - i\epsilon}
\ee
\noindent
The denominator can be written as
\be
\frac{1}{ \nu -\w_r -i\epsilon} = P \left(\frac{ 1}{\nu -\w_r}\right) +
i \pi \de (\nu - \w_r)
\ee
hence, in the limit where $\epsilon \rightarrow 0$ we have,
\bea
2 \left[ n(\w_r) -1 \right] &=& \frac{P}{\pi i} \int_{-\infty}^{\infty}
\frac{\left[ n(\nu)-1\right] d\nu}{\nu - \w_r} +\left[ n(\w_r) -1\right] \nn \\
\left[ n(\w_r) -1 \right] &=& \frac{P}{i \pi } \int_{-\infty}^{\infty}
\frac{\left[ n(\nu)-1\right](\nu + \w_r)} {(\nu - \w_r)( \nu + \w_r)} d\nu
\eea

\noindent
Using our symmetry relation Eq.(\ref{sym}) we find $n_r(\w)$ is an even
function and $n_i(\w)$ is an odd function. We ignore the odd integrands
which give zero upon integration. Thus we can write the real and
imaginary parts of $n(\w)$ where the frequency $\w_r = \w$ is now purely real,
as
\bea
\mbox{Re}\left[ n(\w) \right] &=& 1 + \frac{2 P}{ \pi} \int_0^{\infty}
\frac{ \nu \mbox{Im}\left[ n(\nu) \right] d\nu}{ \nu^2 - \w^2 } \nn \\
&=&  1 + \frac{c P}{ \pi} \int_0^{\infty}
\frac{\alpha(\nu) d\nu}{ \nu^2 - \w^2 } \label{kk} \\
\mbox{Im}\left[ n(\w) \right] &=& -\frac{ 2P}{\pi} \int_0^{\infty}
\frac{ \w \mbox{Re}\left[ n(\nu) -1 \right] d\nu}{ \nu^2 -\w^2 }
\eea

\noindent
where $\alpha$ is the absorption coefficient defined earlier.
These results are consistent with the findings of the text books
\cite{jack,sak,panof} and also Nussenzveig \cite{nuss}, Toll \cite{toll},
Landau and Lifshitz \cite{stat} and Rauch and Rohrlich \cite{rohr}.

\section{Discussion and Conclusions}

We have shown in section 2, that although the Scharnhorst effect suggested
that faster--than--c signals were in principle possible, in practice it
would be impossible to detect any such increase in c. We further suggested
that the original calculations of the effect may be flawed, and that
the effect implies a violation of SR. Later sections of this paper relate
to the case of physical media, by which we mean non--vacuum,
finite energy media in non--curved spacetime. We show that physically
a result $n(\w) \rightarrow 1$ as $\w \rightarrow \infty$ is the only
reliable result based on electron energies being less than infinity.\\

Historically, the result of Eq.(\ref{kk}) was first derived by
Kronig (1926) \cite{kronig}
and the equivalent result for the dielectric constant  was treated by
Kramers in (1927) \cite{kramer}. Kronig was interested in the physical
model behind the derivation of the refractive index and
how many atoms are required before you can sensibly use the
refractive index idea for a bulk material.
Kramers first employed Cauchy integrals to derive the above
dispersion relations. However, neither paper puts any emphasis on
causality.  This comes in much later by Kronig (1942) \cite{kronig2}.
More recent tutorial accounts of the dispersion relations which
include a discussion on causality
can be found, for example see \cite{peter,wolf}.
Text book accounts \cite{jack,sak,panof},
will mention that $n(\w) \rightarrow 1$
as $\w \rightarrow \infty$ but will only argue based on the physical
model for the refractive index. We have shown above that if
$\left[ n(\w) -1 \right]$
satisfies the dispersion relations then, regardless of how it was
derived, Titchmarsh's theorem proves that $\left[ n(\w) -1 \right]$
obeys causality and must also therefore have the limiting result
that $n(\w) \rightarrow 1$ for $\w \rightarrow \infty$ by condition (ii) of
Titchmarsh's theorem. This, to the authors knowledge, has not been stressed
in the literature. No formal proof of $n(\w) \rightarrow 1$ as
$\w \rightarrow \infty$ has been previously given, it has always been tacitly
assumed from a physical model. I would like to stress that the causality
condition (or the dispersion relations) requires this to be true
regardless of the model in use. No model for refractive index can claim to
have a different limiting value on $n(\w)$ as $\w \rightarrow \infty$
without violating causality for $n(\w) -1$.

\noindent
Landau and Lifshitz \cite{ll}, have a footnote in their book relating to
the dispersion relation for the dielectric contant $\epsilon$.

\begin{quote}
``The property $\epsilon \rightarrow 1$ as $\w \rightarrow \infty$ is not
important: if the limit $\epsilon(\infty)$ were other than unity, we should
simply take $\epsilon(\w) - \epsilon(\infty)$ in place
of $\epsilon(\w) -1$ ..."
\end{quote}

\noindent
The same argument would apply to $n(\w) -1$, since it is
$n(\w) -1 \rightarrow 0$ as $\w \rightarrow \infty$ which is implied by
Titchmarsh's theorem to be a condition for the dispersion relation to hold.
If in a case when $n(\infty) \neq 1$ we used instead
$n(\w) -n(\infty)$ then you would regain a valid dispersion relation in a
sense that effect would not precede cause.
It would be necessary to replace
$n(\w) -1$ by $n(\w) -n(\infty)$ everywhere in the dispersion relations,
which is equivalent to a dispersion relation with one subtraction at
$n(\infty)$, to be discussed below.
Landau and Lifshitz were clearly not conserned with relativistic causality.
The dispersion relations require primitive causality only, effect cannot precede
cause.
The author believes the above footnote to be somewhat misleading, especially
now when physicists are conserned with signals traveling at faster--than--c
velocity. The value $n(\infty)=1$ is certainly important.
It is required that $n(\infty) =1$ to show that a front velocity
of a signal propagates at $c/1$ and not $c/n(\infty)$.
Pulse propagation is governed by Maxwell's equations which are certainly
relativistically causal.
Recent articles \cite{junk}, on superluminal signal propagation
based on unphysical models which allow $n(\w) \rightarrow \beta$ when
$\w \rightarrow \infty$ where $\beta < 1$ must therefore be dismissed as
a violation of relativistically causal behavior.

\noindent
Finally, we would like to mention a paper by Weaver and Pao, \cite{pao}.
They derive dispersion relations (for the wavenumber k(\w)) for a general class
of linear homogeneous or inhomogeneous media. The proof proceeds without a
priori knowledge of the dispersion equation for the medium, or rather
without knowledge of the phase velocity $c_{\infty} = \mbox{lim } (\w/k(\w))$
as $\w \rightarrow \infty$. They represent $k(\w)$ as a Herglotz function.
A Herglotz function $k(\w)$ has the following properties;
$k(\w)$ is regular in $I_+$ and Im $k(\w) \geq 0$, (see
Appendix C page 393 in Nussenzveig's book \cite{nuss}).
The Herglotz function has the
desirable feature that its imaginary part is non--negative. The medium is
assumed to be passive, no energy can be given to
the wave ($|G(\w)| \leq 1$) which unfortunately negates the possibility
of an amplifying medium, which may also be causal although its square
integrability may be harder to ascertain. The author finds the use of
Herglotz functions unnecessary (although mathematically valid). The required
proof stems from Titchmarsh's theorem 93, Lemma, which Weaver and Pao seem
aware of. Weaver and Pao have also gone to great lengths to
explain the use of subtractions in the dispersion relations, also
explained in detail by Nussenzveig \cite{nuss}. For example,
if the square integrability condition on $G(\w)$ cannot be satisfied, but the
weaker condition that $G(\w)$ is bounded, $|G(\w)|^2 \leq K_0$, can be
satisfied, then we may construct a new function
\be
H(\w) = \frac{ G(\w) - G(\w_0)}{\w - \w_0} \mbox{\hspace{1.0in} Im }\w_0 \geq 0
\ee
where $H(\w)$ is square integrable and has no poles in $I_+$  and hence
satisfies the dispersion relations.
\bea
H(\w) &=& \frac{P}{i\pi} \int_{-\infty}^{\infty} \frac{ H(\nu)}{\nu - \w} d\nu
\mbox{ \hspace{1.0in} real $\w$} \nn \\
G(\w) &=& G(\w_0) + \frac{\w -\w_0}{i \pi} P \int_{-\infty}^{\infty}
\frac{ G(\nu) - G(\w_0)}{\nu -\w_0} \frac{ d\nu}{ \nu - \w}
\eea
where $G(\w_0)$ is called the subtraction constant. Taking the real part
we find,
\be
\mbox{Re } G(\w) = \mbox{ Re } G(\w_0) + \frac{\w -\w_0}{\pi}
P \int_{-\infty}^{\infty} \mbox{Im } \left[
\frac{ G(\nu) - G(\w_0)}{\nu -\w_0} \right] \frac{ d\nu}{ \nu - \w}
\ee
\noindent
This is known as a dispersion relation for $G(\w)$ with one subtraction. Often,
subtractions will occur at $\w_0 = 0$ or $\w_0 = \infty$.
More than one subtraction is allowed. Using the $\w_0 = \infty$ subtraction
for $n(\w) -1$ we find,
\bea
\mbox{Re } n(\w) &=& \mbox{Re }n(\infty) + \lim_{\w_0 \rightarrow \infty}
\left[ \frac{ \w -\w_0}{\pi} P \int_{-\infty}^{\infty} \mbox{ Im } \left[
\frac{ n(\nu) - n(\w_0) }{\nu - \w_0} \right] \frac{ d\nu}{\nu - \w} \right]
\nn \\
&=& \mbox{Re }n(\infty) +
\frac{P}{\pi} \int_{-\infty}^{\infty} \mbox{ Im }
\left[ n(\nu) - n(\infty) \right] \frac{ d\nu}{\nu - \w} \nn \\
&=& \mbox{Re } n(\infty) + \frac{2P}{\pi} \int_0^{\infty}
\left[ \nu \mbox{ Im }n(\nu) - \w \mbox{ Im }n(\infty) \right]
\frac{d\nu}{\nu^2 -\w^2}
\label{lasteqn}
\eea
\noindent
where we have used the fact that Im $n(\nu)$ is odd.
This is the same as the result implied by Landau and Lifshitz
\cite{stat,ll}, but it only requires primitive causality. For relativistic
causality we also require that $n(\infty)=1$ so that the front velocity of
a signal travels at $c/1$ and not some arbitrary $c/n(\infty)$.\\

We have not commented on situations which relate to highly curved
spacetime. These extreme cases of general relativity may allow for
faster--than--c signals, but we have not studied those situations
here and will not confirm or deny any claims made in the literature. \\

\section*{Appendix}

\noindent
In this section we would like to show that a causal transform of the form
\be
G(\w) = \int_0^{\infty} g(\tau) e^{i\w \tau} d\tau
\mbox{\hspace{0.5in} ( $g(\tau) =0$ when $\tau < 0$ )}
\ee
which is square integrable along the real axis is square integrable along
any line parallel to the real axis in the upper half of the complex plane,
$I_+$.  This condition further implies that
\be
\lim_{x \rightarrow \pm \infty} G(x +iy) =0 \mbox{ \hspace{0.5in} $(y \geq 0)$}
\ee
\noindent
The following mathematical appendix is a summary from Titchmarsh's book
\cite{titch}, Theorem's 93 with Lemma and Theorem 95 which is related.
These arguments can also be found in the book by Nussenzveig
\cite{nuss}, page 22.

\noindent
Consider the analytic continuation of $G(\w)$ into $I_+$. Let $\nu = x+iy$ and
let $\w = x_0 + iy_0$ and consider the following Cauchy integral,
\be
G(\w) = \frac{1}{ 2 \pi i} \int_{\Gamma} \frac{ G(\nu) d\nu}{\nu -\w}
\ee
\noindent
We take the contour $\Gamma$ to be a rectangle with corners at $\pm R$ and
$\pm R + iS$, where both $R$ and $S$ shall tend to infinity.
As we shall show, the integrals of the vertical sides of the contour are both
positive and finite and do not cancel out. They must therefore both be zero.
The magnitude
of the line integral along the right hand side of the contour is given by,
\bea
\left| \int_0^S \frac{ G(R+iy) dy}{R +iy -x_0 -iy_0 } \right|
&\leq& \max_{0 \leq y \leq S} | G(R+iy)| \int_0^S \frac{dy}{\left[
(R-x_0)^2 + (y-y_0)^2 \right]^{1/2} } \nn \\
&\leq& \max_{0 \leq y \leq S} | G(R+iy)|
\ln \left| \frac{ (S-y_0) +\left[ (R-x_0)^2 +(S -y_0)^2 \right]^{1/2} }{
\left[ (R-x_0)^2 +y_0^2 \right]^{1/2} - y_0 } \right| \nn \\
&&
\eea

\noindent
As $R \rightarrow \infty$ and $S \rightarrow \infty$ this implies that
\be
\max_{0 \leq y \leq S} | G(R+iy)| \rightarrow 0
\label{lim}
\ee
since the natural log term tends to $\ln( 1 + \sqrt 2)$ as both $R$ and $S$
tend to infinity.
Similarly the magnitude of the integral on the left hand side of the contour is,

\bea
\left| -\int_0^S \frac{ G(-R+iy) dy}{-R +iy -x_0 -iy_0 } \right|
&=& \left| \int_0^S \frac{ G(-R+iy) dy}{(R +x_0 )-i(y-y_0 ) } \right| \nn \\
&\leq& \max_{0 \leq y \leq S} | G(-R+iy)| \int_0^S \frac{dy}{\left[
(R+x_0)^2 + (y-y_0)^2 \right]^{1/2} } \nn \\
&\leq& \max_{0 \leq y \leq S} | G(-R+iy)|
\ln \left| \frac{ (S-y_0) +\left[ (R+x_0)^2 +(S -y_0)^2 \right]^{1/2} }{
\left[ (R+x_0)^2 +y_0^2 \right]^{1/2} - y_0 } \right| \nn \\
&&
\eea
The natural log has the same limit as above in the case when $R$ and
$S \rightarrow \infty$.  This leads to the same result as Eq.(\ref{lim}),
now written for the limit $R \rightarrow - \infty$. Hence,
\be
\lim_{R \rightarrow \pm \infty} | G(R+iy)| \rightarrow 0
\mbox{ \hspace{0.5in} ($y \geq 0$) }
\ee
\noindent
What remains of the contour integral are the bottom and top line integrals
which can be written, for $R \rightarrow \infty$, as;
\be
G(\w) = \frac{1}{2 \pi i} \left[
\int_{-\infty}^{\infty} \frac{G(x) dx}{x -x_0 -iy_0} -
\int_{-\infty}^{\infty} \frac{G(x+iS)  dx}{x +iS -x_0 -iy_0} \right] \;\; .
\ee
\noindent
According to Schwarz's inequality,
\bea
\left| \int_{-\infty}^{\infty} \frac{G(x+iS) dx}{ x+ iS -x_0 -iy_0} \right|^2
&\leq& \int_{-\infty}^{\infty} |G(x+iS)|^2 dx
\int_{-\infty}^{\infty} \frac{dx}{(x-x_0)^2 +(y-y_0)^2} \nn \\
&\leq& \frac{ \pi K}{S -y_0}
\eea
\noindent
where we have used Titchmarsh's theorem, condition (ii) in the last step.
As $S \rightarrow \infty$ we see that the above expression tends to zero.
Hence,
\be
G(\w) = \frac{1}{2 \pi i}\int_{-\infty}^{\infty} \frac{G(\nu)d\nu}{\nu-\w}
\mbox{\hspace{0.5in} (Im $\w \geq0$)}
\ee
\noindent
which expresses the value of $G(\w)$ at any point in the upper half
of the complex plane in terms of values on the real axis.

%\begin{acknowledgments}
\section*{Acknowledgments}

\noindent
I would like to thank the members of T4 theoretical division,
Los Alamos National Laboratory, for their
support during a difference--in--pay leave from California State
University Fullerton, during which time this work was completed.
In particular I would like to thank
Dr Peter Milonni who suggested the problem and for helpful
discussions and Dr. Daniel James for financial
support during my leave. An extended version of this work has been written by
Mr. Robert H. Gibb for his Masters thesis at CSUF physics department
August 2003.
%\end{acknowledgments}

%%\begin{noteinproof}
%%\end{noteinproof}

\end{document}